# Defect Engineering in Large-Scale CVD-Grown Hexagonal Boron Nitride: Formation, Spectroscopy, and Spin Relaxation Dynamics

*Ivan V. Vlassiouk[1]\*, Yueh-Chun Wu[2], Alexander Puretzky[1], Liangbo Liang[1], John Lasseter[1], Bogdan Dryzhakov,[1] Ian Gallagher[2], Sujoy Ghosh[1], Nickolay Lavrik[1], Ondrej Dyck,[1] Andrew R. Lupini,[1] Marti Checa,[1] Liam Collins,[1] Huan Zhao,[1] Farzana Likhi,[3] Kai Xiao,[1] Ilia Ivanov,[1] David Glasgow,[4] Alexander Tselev,[5] Benjamin Lawrie,[1,2] Sergei Smirnov[6] and Steven Randolph[1]*

[1]Center for Nanophase Materials Sciences, Oak Ridge National Laboratory, Oak Ridge, TN, USA.

[2]Materials Science and Technology Division, Oak Ridge National Laboratory, Oak Ridge, TN, USA

[3]Materials Science and Engineering, University of Houston, Houston, TX, United States

[4]Chemical Sciences Division, Oak Ridge National Laboratory, Oak Ridge, TN, USA

[5]Aveiro Institute of Materials and Department of Physics, University of Aveiro, Aveiro, Portugal

[6]Department of Chemistry and Biochemistry, New Mexico State University, NM, USA.

Email: vlassioukiv@ornl.gov

**Abstract.**

Recently, numerous techniques have been reported for generating optically active defects in exfoliated hexagonal boron nitride (hBN), which hold transformative potential for quantum photonic devices. However, achieving on-demand generation of desirable defect types in scalable hBN films remains a significant challenge. Here, we demonstrate that formation of negative boron vacancy defects, $V_B^-$, in suspended, large-area CVD-grown hBN is strongly dependent on the type of bombarding particles (ions, neutrons, and electrons) and irradiation conditions. In contrast to suspended hBN, defect formation in substrate-supported hBN is more complex due to the uncontrollable generation of secondary particles from the substrate, and the outcome strongly depends on the thickness of the hBN. We identify different defect types by correlating spectroscopic and optically detected magnetic resonance features, distinguishing boron vacancies (formed by light ions and neutrons) from other optically active defects emitting at 650 nm assigned to anti-site nitrogen vacancy ($N_BV_N$) and reveal the presence of additional "dark" paramagnetic defects that influence spin-lattice relaxation time ($T_1$) and zero-field splitting parameters, all of which strongly depend on the defect density. These results underscore the potential for precisely engineered defect formation in large-scale CVD-grown hBN, paving the way for the scalable fabrication of quantum photonic devices.

**Introduction.**

Optically active defects in various solid-state materials that can be treated as two-level quantum systems have become a versatile platform for developing emerging quantum technologies.[1, 2] The recent surge in the exploration of such defects in two-dimensional (2D) materials has been driven by their unique properties and, in part, by a wide variety of such van der Waals materials that can enable a plethora of functionalities.[3] It is anticipated that such versatile 2D heterostructures should allow the development of integrated quantum photonic systems tuned to specific applications. Such defects have been studied in many 2D materials primarily focusing on transition metal chalcogenides (TMDs)[4] and hexagonal boron nitride (hBN).[5, 6] The latter recently received particular attention due to its wide-bandgap (6 eV) where defects exhibiting single-photon emitter (SPE) properties in a wide range of emission energies have been demonstrated.[7-9]

Defects in hBN with broadband emission near 800 nm were recently assigned to negatively charged boron vacancy ($V_B^-$), which has a spin-triplet ground state,[10] while the origin of 650 nm emission is still debated with some reports assigning it to anti-site nitrogen vacancy ($N_BV_N$),[11] or carbon containing defects.[12] The first report on paramagnetic defects created in pyrolytic boron nitride by neutron bombardment is over three decades old,[13] but only recently it has regained attention.[10, 14-17] Such $V_B^-$ defects exhibit spin-dependent optical transitions and have long relaxation times, making them attractive candidates for developing quantum optical devices in various applications, including quantum sensing and communications.[18, 19] It was shown that boron vacancies can be formed not only by neutron bombardment but also by electrons, and ions of different type.[20]

Most of the previously reported experiments on defect introduction were performed on either high-quality bulk hBN crystals or on exfoliated sub-millimeter-sized samples.[3] This is a result of the need to start with defect-free samples, and technologically relevant CVD (chemical vapor deposition)-grown hBN films have lacked that quality and scale. Recently, we[21] and others[22-24] have demonstrated a facile but robust method of high-quality hBN film deposition on a large-scale using molecular nitrogen and various solid boron sources with an extremely low density of defects achieved, comparable with that of bulk hBN crystals. Thickness control of hBN films from monolayer to >100 nm was demonstrated by adjusting catalyst composition.[21] This allows for the synthesis of high quality hBN films suitable for the rational design of defects with desired properties.

Here, we investigate the formation of optically active defects in large-scale CVD grown hBN films using different types of particle bombardment (ions, neutrons, and electrons) and demonstrate the selective formation of specific defect types. We focus on the 800 nm emission (preferentially formed by light ions and neutrons) attributed to $V_B^-$ defects, that competes with defects of different origins emitting near 650 nm (mostly induced by heavy ions) assigned to anti-site nitrogen vacancy ($N_BV_N$). The thickness of hBN plays a critical role in defect formation, and the outcomes differ between suspended hBN films and those on substrates. This distinction arises from poor control over the type and energy/spatial distributions of secondary particles generated upon collision with the underlying substrate which leads to the formation of mixed defects. We use cathodoluminescence (CL), photoluminescence (PL), Raman spectroscopy, Kelvin probe microscopy, and optically detected magnetic resonance (ODMR) for deeper characterization and understanding of the formation mechanisms of these defects. Using ODMR, we confirm $V_B^-$ ground triplet state, and we observe a strong correlation between the spin relaxation times ($T_1$) and the paramagnetic defect densities that influence $T_1$ through a cross-relaxation mechanism. Our findings highlight the importance of balancing photoluminescence brightness and $T_1$, both of which strongly depend on the defect density. $T_1$ depends on the environment: at room temperature it decreases from 15 μs at low defect density with weak dependence on the external magnetic field to less than 1 μs at high defect density and strong dependence on the field.

The general conclusion is that in high quality CVD-grown hBN films, $V_B^-$ defects, among others, can be controllably produced yielding a significant potential for large-scale quantum photonic devices.

Nevertheless, reproducible device fabrication and optimized performance on a large scale would ultimately require ultimate fine tuning of synthesis to achieve a uniform hBN thickness on the wafer scale.

**Results and discussion.**

The penetration depth for a particle into hBN depends on multiple factors including the particle mass, kinetic energy, and angle of incidence, all of which need to be taken into account when generating the defects. If the incident particles penetrate through the hBN layer, they will interact with the substrate, scatter, and possibly backscatter or generate other particles. It introduces complexity into defect generation mechanisms and their spatial localization. For example, bombardment of $Si/SiO_2$ - commonly used as substrates in many hBN experiments - by $He^+$ ions leads to strong photoluminescence at ~700 nm, which sometimes mistakenly is attributed to hBN (**SI, Fig S1**). The $He^+$ beam at 30 keV used in this study has a penetration depth or end-of-range (EOR) length of ~240 nm.[15] Importantly, due to the synthesis conditions, most of the hBN films used in this study are thinner than 250 nm (**SI, Figs S2-S5**). Thus, to decouple the effects of the substrate during the bombardment, we compare defect formation in suspended hBN membranes and hBN films on bulk substrates.

*1. Suspended hBN membranes.* As shown in **Fig. 1a**, CVD-grown hBN films are transferred onto $Si/SiN_x/Au$ membranes and thus have regions that are suspended and on top of gold. The gold layer was deposited on top of $SiN_x$ to screen the strong 700 nm $SiN_x$ fluorescence that makes the spectroscopic characterization of hBN very hard (**SI, Fig. S6**) and simulate the appearance of hBN on a microwave (MW) waveguide used later in the ODMR study. The optical microscope image highlights different colors arising from the thickness nonuniformity of the hBN film. Most of the hBN is ~90-120 nm thick and has a goldish color, while the islands are ~240 nm thick and have a pinkish color. The thickness of hBN was determined by optical reflectance spectra and confirmed by (destructive) plasma-focused ion beam (PFIB) etching coupled with atomic force microscopy (AFM) measurements (**SI, Fig. S2-S5**). For hBN suspended over large holes, the film's color also varies due to the thickness variation, which can be monitored by reflectance or transmittance spectra. These nonuniformities in the thickness can be taken into account in data analysis, as explained later.

A typical Raman map of suspended CVD hBN is shown in **Fig. 1b** along with the histogram of the $E_{2g}$ band over the whole measured area. The $E_{2g}$ band corresponds to the in-plane vibration of $sp^2$ hBN at 1365 $cm^{-1}$ and is considered a primary metric of the hBN quality and uniformity. The map reveals weaker Raman intensity for suspended hBN, at least for ~120 nm thick samples. This can be rationalized by an increased reflectance from the gold. At the excitation wavelength (532 nm), the reflection from gold in air is ~0.7, but it is different at the interface with the hBN film and is strongly depends on hBN thickness (**SI, Fig. S2**). Our typical hBN samples on gold show approximately twice the Raman $E_{2g}$ intensity compared with suspended regions, (**SI, Fig. S7**) while some samples of other thicknesses can show an increase as much as fivefold possibly due to plasmon excitations in Au. The FWHM of the $E_{2g}$ band in the histogram has a mean value of ~8.5 $cm^{-1}$, only slightly larger than the 6-8 $cm^{-1}$ FWHM for typical exfoliated samples.

After bombardment of hBN with ions, two additional Raman bands appear; we denote them *E'*, at ~1290 $cm^{-1}$, and *D*, at 450 $cm^{-1}$. **Fig.1c** shows the case of irradiation with 30 keV $He^+$ ions. The main $E_{2g}$ Raman line also significantly broadens with increased defect density. While *E'* was previously reported for exfoliated samples after irradiation,[15] the *D* band was identified only for bulk crystals.[25] Both these bands can also be observed in the anti-Stokes part of the Raman signal (**SI, Fig. S8**). The negatively charged $V_B^-$ defect has $D_{3h}$ symmetry and is predicted to have ~1300 $cm^{-1}$ and ~325 $cm^{-1}$ vibration modes[26] that closely correlate with the observed positions of the *E'* and *D* lines. Therefore, we attribute these two modes to $V_B^-$. Additional evidence supporting the assignment of these bands to $V_B^-$ is presented in the following sections.

The introduction of defects is accompanied by the appearance of strong photoluminescence (PL) with a maximum at 800-900 nm (**Fig. 1c,** center) and cathodoluminescence (CL). In the latter, an additional 650 nm emission peak is also observed; we label these bands G1 and G2, respectively (**Fig. 1c,** right). A sharp feature at ~575 nm corresponds to the Raman signal from hBN. As mentioned above, there is no PL on most areas in hBN prior to irradiation with only rare SPE-like hotspots indicating a very high quality of our "as synthesized" hBN (**SI, Fig. S9**). Similarly, in CL "as synthesized" hBN has much weaker emission in the UV-vis spectrum due to defect-related bands[27] that are likely formed during electron bombardment[28-32] as supported by the results presented in the next sections. Stronger emission further in the UV has been reported in the literature[33] but due to our CL detector is limited to operation at 350-1000 nm.

Focused ion beams can pattern such defects in a desired pattern. **Fig. 1d** shows images of the area patterned with a 25 $He^+/nm^2$ dose recorded by three different techniques: Kelvin probe force microscopy (KPFM), CL, and PL. Importantly, AFM topography does not resolve the patterned region, while KPFM clearly contrasts the bombarded regions. Lower measured potential on the irradiated areas likely arises from increased conductivity due to embedded $He^+$ ions and newly formed charged defects and their counterions.

**Figure 2a** shows PL and Raman maps of a suspended hBN membrane bombarded with 50 $He^+/nm^2$ dose at 30 keV. This is the only energy used here for $He^+$ ions. The maps show integrated intensities over the two spectral ranges, 700-900 nm (PL) and 565-580 nm, the latter of which corresponds to the combined $E'$ and $E_{2g}$ Raman signals. Note a strong fluorescence background signal from the exposed circular $SiN_x$ membrane edge uncovered by gold (**SI, Fig. S6**). Regions with high/low intensities coincide for PL and Raman maps, suggesting that it is due to the hBN thickness variation - thicker hBN regions produce stronger PL and Raman. The PL/Raman ratio map gives a constant value, independent of hBN thickness, indicating that the structure/density of the defects has only a weak dependence on the kinetic energy of $He^+$ ions as they traverse the entire hBN thickness and dissipate their energy. Indeed, the observed ratio evens out the original maps' nonuniformities and demonstrates a narrow ratio distribution (**Fig 2a**, right). Thus, PL normalization by the Raman intensity provides a reasonable metric for estimating $V_B^-$ yield under different bombardment parameters for the thicknesses below EOR.

**Figures 2b-c** compare changes in the Raman and PL spectra for $He^+$ and $Ar^+$ bombardment (left and right, respectively) using the same 30 keV ion energies. Two obvious differences can be recognized. First of all, lighter $He^+$ ions induce larger $D$ and $E'$ bands in the Raman spectra, while for heavier $Ar^+$ ions, these peaks have much lower intensities and are present only at low dosages. Second, the PL shows a correlated behavior with Raman – the 800 nm peak is observed for $He^+$ at all dosages but only at low dosages for $Ar^+$, for which at large dosages another, 650 nm band overwhelms the 800 nm band. **Figure 3** summarizes these observations with finer details.

Heavy ions have more efficient energy transfer to boron and nitrogen and thus are more effective in forming complex defects with knocked out atoms. Such a trend was previously observed for graphene.[34] **Figure 3a** illustrates how the defects, evaluated as the $E'/E_{2g}$ intensity ratio, evolve as a function of dosages for ions of different mass: $He^+$ (black), $Ne^+$ (red), $Ar^+$ (blue), and neutrons (magenta). For the lightest ion $He^+$, much higher doses are required for the $E'$ feature to appear, which eventually reaches a maximum at ~ 100 $ions/nm^2$ and declines. The heavier $Ne^+$ ions produce defects at lower dosages. The heaviest $Ar^+$ ions succeed in making such defects at even lower dosages - the dosage of a maximum for $E'/E_{2g}$ ratio for $Ar^+$ is roughly two orders of magnitude lower than for $He^+$. At the same time, the maximum $E'$ intensity for $Ar^+$ irradiated hBN is substantially lower compared to $Ne^+$ and especially $He^+$. These results indicate that the $E'$ band corresponds to defects produced with greater yield under lighter ion bombardment.

The trend agrees with the expected decline of the energy transfer efficiency for particles of small masses but further lowering the mass of irradiating particles (by moving to neutron or electron irradiation, for example) is counterproductive. A declining cross section for energy transfer not only requires much higher dosages for defect production but also leads to longer EOR and thus a greater effect of interaction with the substrate for thin hBN films. Neutrons as irradiating particles of suspended hBN are closer in performance to light ions even though our source has a broad energy distribution (**SI, Fig. S13**) and there is a route for a thermal neutron transmutation reaction with $^{10}$B leading to the formation of high energy $^{7}$Li, $^{4}$He, and gamma photons (**SI**). The Raman and PL spectra of suspended hBN after neutron bombardment exhibit the characteristic *D* and *E'* Raman modes along with the 800 nm PL band.

The results of keV electron bombardment are more complex. Large electron dosages (~$10^6$ e/nm$^2$) are required to observe 650 nm PL (with weak 800 nm shoulder) and 1580 cm$^{-1}$ Raman on "as-synthesized" suspended hBN, however, possible electron-beam-induced carbon deposition[35] complicates the analysis (**SI, Section 6**). It is not apparent why electrons are so prone to producing such defects. We suspect that it is due to a significant difference in the cross sections for producing initial defects and their 'alteration' by second collision, where the second one is significantly greater. It is apparent from **Fig. S14c** that no recognizable single atom knockout defects are visible under electron irradiation; all defects appear with multiple atoms missing, and the number and size of those defects grows with the dosage. Indeed, these results are consistent with existing literature showing that reducing the electron-beam energy can in fact increase the defect formation efficiency. [36, 37]

Notably, the *E'/D* ratio remains nearly constant (1.68 ± 0.26 for 532 nm excitation), regardless of whether neutrons or different ion species are used and regardless of the dose chosen for defect creation (**Fig. 3b**). This strongly suggests that the *D* band is either a different vibrational mode of the same defect, as was predicted for $V_B^-$,[26] or another defect type that incidentally always forms in the same proportion, which is unlikely. This stands in stark contrast to graphitic materials, where different types of defects also produce two Raman bands identified as a *D* peak at ~1330 cm$^{-1}$ and *D'* at ~1620 cm$^{-1}$ along with their combination peak, *D+D'*, at ~2940 cm$^{-1}$. The irradiation-dependent changes in the ratio of intensities of those peaks, *D/D'*, have allowed for the characterization of different defect types.[38]

As described above, variations in PL intensity at 800 nm for He$^+$ irradiated samples follows the thickness and, when normalized by the $E_{2g}$ intensity, it gives a uniform image (**Fig. 2a**). Similar normalization can be assessed for samples irradiated with other ions as shown in **Fig. 3c**. First, normalization by either $E_{2g}$ (black) or *E'*+$E_{2g}$ integrated intensity (green) gives a similar dependence. Second, the 800 nm PL intensity mirrors the trends of *E'* and *D* (**Figs. 3a,b**) across different doses and ion types, suggesting that the same defects are responsible for both the Raman and 800 nm PL features.

**Figure 3d** shows the PL intensity ratio between 650 nm and 800 nm bands for samples irradiated with different ions. For suspended hBN membranes, a pronounced 650 nm peak appears in the PL spectra only of the Ar$^+$ bombarded samples (**Figs. 2c,3c**), at least for the doses used. Nevertheless, in the CL spectra it can be observed for a He$^+$ bombarded samples too; its intensity increases and the peak position hypsochromically shifts with increasing dose (**SI, Fig. S14**). Notably, this peak is also observed in the PL of He$^+$ bombarded samples taken after CL experiments. (**SI, Fig. S10**), which suggests that its appearance is caused by irradiation with electrons. The 650 nm PL and a ~1580 cm$^{-1}$ Raman band can be attributed to the formation of $N_BV_N$ defects[11] - an anti-site nitrogen vacancy in which a nitrogen atom migrates to a boron vacancy, leaving behind a neighboring nitrogen vacancy, suggesting that e-beam bombardment leads to the transformation of $V_B^-$ (prepared by He$^+$ bombardment) into $N_BV_N$.

To understand the experimental defect-related Raman spectra, we carried out first-principles density functional theory (DFT) calculations of monolayer hBN with a $V_B^-$ defect and a $N_BV_N$ defect, as shown in **Figure 4**. Raman spectra are found to be strongly dependent on the defect structure. In **Fig. 4a**, a Raman peak at ~389 cm$^{-1}$ is dominant in the spectrum, which corresponds to a characteristic phonon mode closely related to the $V_B^-$ defect as its vibrations are largely localized around the vacancy site. This is consistent with a prior work predicting such a defect phonon mode at ~325 cm$^{-1}$.[26] In contrast, for the $N_BV_N$ defect (**Fig. 4b**), the strongest Raman peak is around 1540 cm$^{-1}$ with a vibration pattern more or less distributed across the whole structure. Such defect-dependent Raman features corroborate the experimental observations that the $V_B^-$ defect yields a strong Raman defect peak around 450 cm$^{-1}$, while the $N_B V_N$ defect leads to a strong Raman band around 1580 cm$^{-1}$. Our calculations slightly underestimate the frequencies of Raman peaks, but this does not affect the conclusion.

We attempted to resolve the 650 nm PL hBN defect structure using STEM in a few-layer hBN sample that served as a model for our multilayer hBN samples. In monolayer hBN, the distinct contrast between N and B atoms makes it possible to resolve the atomic structure of the defects with practically unambiguous element assignment.[39] However, in a multilayer AA' stacking, where B and N atoms are on top of each other, it is next to impossible to distinguish between even the simplest defects, such as $V_B$ and $V_N$, relying exclusively on the contrast. Moreover, we found it difficult to conclusively determine whether the observed defects by STEM were induced by prior ion bombardment or generated under the electron beam during imaging. In fact, multiple complex defects are formed even during bilayer hBN imaging, and pre-existing defects - possibly introduced by ion beams[40] - enlarged under the STEM beam after ~$10^8$ e/nm$^2$ dosage (**SI, Fig. S16**).

*2. hBN on substrates.* As previously mentioned, the types and yields of defects in hBN on a substrate can differ from those in suspended samples for the same type of irradiation, especially when the hBN thickness, $L$, is lower than the EOR of the incident particles. When an incident beam fully penetrates the hBN overlayer ($L$ < EOR), its interaction with the underlying substrate inevitably introduces additional defects in the overlying hBN. Even when the hBN layer exceeds the ions' penetration depth ($L$ > EOR), a substantial decline of the ion's kinetic energy by the end of EOR can lead to diverse defect generation mechanisms and variations in the defect yields.

Light particles have a longer EOR and thus this effect is more pronounced for them. The effect of the substrate is illustrated in **Fig. 5**. First, we consider neutron irradiation with broad energy distributions (**SI, Section 5**) and a very large EOR, allowing the neutron beam to penetrate hBN of all thicknesses used in this study and interact with the Au substrate. The Raman and PL spectra of suspended hBN after neutron bombardment exhibit the characteristic *D* and *E'* Raman modes along with 800 nm PL, similar to the effects observed when hBN is bombarded with light particles, indicating formation of $V_B^-$ defects (**Fig.5a**).

However, the situation changes drastically for supported hBN. A prominent 1580 cm$^{-1}$ Raman band appears, along with pronounced 650 nm PL, previously observed in suspended hBN after e-beam (**Fig. S16**) and Ar$^+$ (**Fig. 2**) bombardment at high dosages.

**Figure 5b** shows Raman and PL spectra of hBN on a gold/sapphire substrate that has been irradiated with 50 He$^+$/nm$^2$. To highlight how the variations in hBN thickness change the outcome, we intentionally deposited hBN with a significant thickness nonuniformity. The gold stripe serves as a microwave waveguide for ODMR experiments described in the following section. The Raman spectra show that thinner regions (~90 nm) exhibit a significantly enhanced E' peak, with $E'/E_{2g}$ ratio reaching 2.5. Additionally, the PL signal at 800 nm, normalized by the Raman intensity, reaches values around 30 – significantly greater than those observed in suspended samples (**Fig. 3a,c**). Also note the appearance of a broad shoulder around ~1580 cm$^{-1}$ that was absent in the suspended samples under similar conditions. He$^+$ ions upon reaching gold underneath hBN inelastically scatter and produce secondary electron emission.[41] The latter are responsible for the 1580 cm$^{-1}$ shoulder in Raman and 650 nm emission on e-beam bombarded samples besides the 'normally' produced $V_B^-$ defects, as for suspended samples.

Interestingly, some samples exhibit hBN "waves", formed unintentionally during the transfer process, that exhibit a strong 800 nm PL band even with Ar$^+$ bombardment, which was never observed for suspended or flat samples on gold substrate (**SI, Fig. S11**). A similar effect was observed on neutron bombarded samples too. We do not attribute this effect to variations in the effective bombardment angle due to the

"waviness" of the hBN film, as no significant dependence was observed when varying the Ar+ beam incident angle from 90° to 30°. Nevertheless, it is clear that the greater signal is not only due to the interference contrast either, as a greater density of defects is clearly identifiable by a greater $E'/E_{2g}$ for the 'wavy' part. Mechanical stress/strain[42, 43] in "wavy" hBN can be one alternative explanation for the enhanced quantum yield of defects but it does not seem to be sufficient either. This illustrates additional complexity and highlights the importance of substrate effects, which were well controlled in our experiments.

3. **Optically detected magnetic resonance.** We will focus on the $V_B^-$ defects emitting at 800 nm as the most promising defect type. The assignment of these defects to $V_B^-$ has been proposed previously[10] and we confirm it here. Besides, we were able to obtain additional important previously unknown information, as described below.

A simplified $V_B^-$ energy-level diagram highlighting major transitions is shown in **Fig. 6a.** The triplet ground state (GS) of the $V_B^-$ defect comprises three spin sublevels with $m_s = 0$ separated from $m_s = \pm 1$ by zero-field splitting (ZFS) characterized by parameter $D$. The excited state (ES) is also a triplet (S=1) while a metastable state (MS) is a singlet (S=0).[10, 16, 17] Non-radiative intersystem crossing (ISC) transitions between S=1 and S=0 states are indicated by blue arrows. Excitation from the GS to ES is shown by green arrows, while radiative PL emission from the ES to GS is represented by red arrows. Notably, the ISC transition from ES (S=1) to MS (S=0) is faster from $m_s = \pm 1$ ($|\pm 1\rangle$) than $m_s = 0$ ($|0\rangle$).[44] This spin-dependent ISC rate enables optical detection of spin transitions, a key feature for quantum applications.

For optical detection of magnetic resonance (ODMR) a sample needs to be exposed to a microwave (MW) field, for which gold coplanar waveguides fabricated on sapphire substrates are used. A sample of hBN is placed on such a waveguide and irradiated directly on it. An optical microscope image of a device geometry, along with a schematic of the $V_B^-$ defect energy diagram, are shown in **Fig 6a**. The MW magnetic field ($B_1$) is oriented along the gold surface, while the applied static magnetic field ($B_0$) is perpendicular to the waveguide plane, *i.e.,* $B_1 \perp B_o$. Under continuous-wave (CW) laser excitation, when the MW is off-resonance, the GS is more spin-polarized in the $|0\rangle$ state, resulting in maximum PL. However, when the MW field drives transitions to $|\pm 1\rangle$, the PL intensity decreases due to the enhanced quenching caused by ISC.[10] The transitions can be induced by MW excitation not only in the GS but in the ES as well but at different frequency.[16]

Neglecting hyperfine interactions, the spin Hamiltonian is given by eq. (1), with the Z-axis oriented perpendicular to the hBN plane:

$$H = D\left(S_z^2 - \frac{S(S+1)}{3}\right) + E(S_x^2 - S_y^2) + g\mu_B \boldsymbol{BS} \qquad (1)$$

Here $E$ and $D$ are the transverse and axial ZFS parameters aligned with and perpendicular to the symmetry axis of the $V_B^-$ center, respectfully. S is the electron spin (S=1), $B_0$ is the external magnetic field, $S_{x,y,z}$ are the spin operators, g is the electron Lande g-factor, and $\mu_B$ is the Bohr magneton.

**Fig. 6b** shows ODMR spectrum at zero external magnetic field ($B_0 = 0$) using CW excitation for a sample after He+ irradiation detected at the 800 nm band corresponding to $V_B^-$ defects. In the ground state, the splitting between $m_s = 0$ and $m_s = \pm 1$ is $\nu_0 = D/h \sim 3.5$ GHz. The splitting between $m_s = 1$ and $m_s = -1$ sublevels is also visible. It is given by E/h at zero field and increases with applied $B_0$ as:

$$\nu_{1,2} = \nu_0 \pm \frac{1}{h}\sqrt{E^2 + (g\mu_B B_0)^2} \qquad (2)$$

Both $D$ and $E$ are primarily determined by the magnetic dipole-dipole interactions of the spins in the triplet state (GS) and were previously thought to be intrinsic characteristics of the $V_B^-$ center.[10] We have found that the parameter $E$ noticeably depends on the dosage (**Fig. 6c**) while $D$ remains unaffected (**SI,**

**Fig. S12**). Similar behavior was previously observed for NV centers in diamond.[45] The value of $E = 51$ MHz at the lowest dose of 0.37 He+/nm² is close to the previously reported for the ground state of $V_B^-$ [10] but increasing irradiation doses lead to its almost two-fold increase at 50 He+/nm². The increase is quite gradual, close to logarithmic. We found the ODMR contrast to be independent on the He⁺ dose, ~3.4 ± 0.7% at room temperature for our bombardment conditions and MW power. To ensure consistency for our ODMR results, we used hBN with a thickness of L~250 nm for all doses, since varying hBN thicknesses can yield different defect densities/types as shown in **Fig 5b**. We also screen all our ODMR samples to ensure that the 650 nm PL and 1580 cm⁻¹ Raman band are absent.

Increasing $B_0$ causes the $m_s=\pm 1$ levels to split further apart in accordance with eq. (2), see **Figs. 6d,e**. This Zeeman splitting corresponds to the Lande g-factor close to that of a free electron, $g_e$=2.00, as was previously reported for $V_B^-$ defects.[10]

The spin-lattice relaxation time ($T_1$) can be measured by a sequence of polarization and delayed readout pulses, as shown in the inset of **Fig. 7a**. The room temperature spin-lattice relaxation time ($T_1$) for $V_B^-$ defects in exfoliated high-quality hBN was measured to be around 18 μs,[14] which is comparable to 15 μs measured in our CVD-grown samples at the irradiation dose of 0.37 He⁺/nm² ($E'/E_{2g}$~0.03). At high defect densities, the reduced distance between defects enhances dipole-dipole interactions between unpaired electron spins, facilitating more efficient energy exchange between the spins and the environment.[46] This leads to shorter $T_1$, which decreases to less than 1 μs at the irradiation dose of 50 He⁺/nm² ($E'/E_{2g}$=1.3)(**Figs. 7a**). The decline has a nearly logarithmic dependence on the dose, similar to that of $E$ dependence. The similarity hints at a possibly common cause. Note that $T_1$ dependence on defect density was previously reported for NV centers in diamond at room temperature,[45] which switched to a much stronger linear variation of $T_1$ with the dose at cryogenic temperatures.[47] The authors proposed different mechanisms of relaxation - interaction with phonons at high (room) temperature and cross relaxation with neighboring spins at low temperature. We believe that interaction with neighboring spins such as other defects and restrained He⁺ ions and knockout atoms/ions could be responsible for relaxation in $V_B^-$ centers even at room temperature. The same defects cause the changes in $E$ as well.

The importance of spin-spin interactions even at room temperature is supported by the weak dependence of $T_1$ on $B_0$ at low defect densities in both supported and suspended hBN **(Fig. 7b**, top**)** and the strong dependence observed at high defect concentrations **(Fig. 7b,** bottom**)**.[46] Cross-relaxation mechanisms are more profound at the points of resonance[44, 48] - where the energy levels of two neighboring spin systems become resonant and induce efficient energy exchange between them. Such points can be observed by varying magnetic field. In the given range up to 400 G, $T_1$ at high defect densities ($E'/E_{2g}$=1.3) initially rises with $B_o$ but declines beyond 200 G. The level anticrossing in $V_B^-$ for the ground state between $m_S$=-1 and $m_S$=0 occurs at $B_0$ ~ 1250 G but between $m_S$=-1 of the ground state and $m_S$ = 1 of excited state (for which $D_e \approx$ 2.1GHz and $E_e$ ~ 74 MHz) the anticrossing occurs near ~200 G.[16] Other dips in PL contrast due to anticrossing with S=1/2 occur near ~400 G and 600 G.[47] A combination of these resonances for $T_1$ at high fields and the rise away from zero field results in an effective maximum of $T_1$ at ~200 G. The dependence is different at low defect densities ($E'/E_{2g}$ ~0.3), where $T_1$ slightly decreases with increasing field.

The anticrossing mechanism, even in samples with low defect density, becomes prominent at low temperatures. **Figure 7c** (top) presents $T_1$ as a function of $B_0$ acquired at T = 80 K, while the ΔPL contrast acquired at the same temperature is shown in **Fig. 7c** (bottom). The levels crossing labels are shown at the top. The most significant decreases in $T_1$ are observed for ½⇔GS ($m_s$=-1) and ES ($m_s$=-1)⇔ ES ($m_s$=0). Importantly, we do not observe a significant difference in the peaks between suspended and supported hBN, suggesting similar densities of paramagnetic defects with S = 1/2. However, $T_1$ for suspended hBN (220 μs) is slightly longer than that of supported hBN (190 μs) at T = 80 K at the same defect densities.

**Conclusions.** We report that specific defect types in large-area CVD-grown hexagonal boron nitride (hBN) can be selectively formed by choosing the bombarding particles. Among the investigated ions of different masses, neutrons, and electrons, He$^+$ ions represent the best choice for preferential creation of negative boron vacancies $V_B^-$ for suspended and substrate-supported hBN with Ne$^+$ and neutrons being the second best. These boron vacancies exhibit two characteristic Raman bands at ~1300 cm$^{-1}$ and ~450 cm$^{-1}$ and emit at 800 nm. Heavier ions (Ar$^+$) and electrons within the tested energy range primarily induce other defects with a 1580 cm$^{-1}$ Raman band and 650 nm photoluminescence (PL), which are likely anti-site nitrogen vacancy ($N_B V_N$) defects.

For hBN supported on a substrate, which is more appropriate for many applications, the outcome of irradiation can significantly differ from that of suspended hBN, particularly when the end-of-range (EOR) of the bombarding particles exceeds the hBN thickness. The secondary particles generated upon collisions with the substrate lead to poorly controlled defects even with He$^+$ ions, although this choice remains the best overall.

The spin-lattice relaxation time, $T_1$, shows a strong correlation with the defect density, decreasing at room temperature from 15 μs at low defect densities to <1 μs at high defect densities following approximately logarithmic dependence. The observed dependence of $T_1$ on the external magnetic field confirms the role of the cross-relaxation mechanism and points to the presence of "dark" paramagnetic defects, such as irradiating ions and knocked out atoms/ions. The in-plane zero-field splitting parameter, $E$, increases with defect density close to logarithmically, while the $D$ parameter remains unchanged. These findings demonstrate the potential for precise engineering of defect formation in large-scale CVD-grown hBN, paving the way for the scalable fabrication of quantum photonic devices, particularly for quantum sensing and communication applications.

**Methods.**

*hBN synthesis and transfer.* A previously reported hBN synthesis approach, utilizing solid boron and molecular nitrogen as precursors, was employed.[21] In brief, a 50 μm thick Invar foil served as a catalytic substrate for growth. The solid boron-containing precursor was placed in a 2x2'' crucible, with the foil positioned on top. A 3-inch hot-wall quartz tube furnace was heated to 1100°C under a 2.5% Ar/$H_2$ mixture. Argon was replaced with nitrogen gas, and the growth process continued for 30 minutes. Subsequently, $FeCl_3$ etchant was used to dissolve the foil. The transfer was conducted without a PMMA supporting overlayer, as the hBN film was sufficiently thick to withstand manipulation during the process.

*Defect introduction.* Zeiss Orion NanoFab Helium Ion Microscope was used for $Ne^+$ and $He^+$, Thermo Scientific Helios 5 Hydra UX DualBeam Multiple Ion Plasma FIB/SEM for $Ar^+$, Zeiss Merlin SEM for electron irradiation, and bombardment by neutrons were done at High Flux Isotope Reactor at Oak Ridge National Laboratory (**SI, Section 5**).

*CL.* The cathodoluminescence (CL) measurements were conducted using an FEI Quattro environmental SEM equipped with a Delmic Sparc CL collection module, which employs a parabolic mirror to collect CL signals from the sample under electron beam excitation. An electron beam at 5 kV and 110 pA was used with a 700 ms exposure time for CL point spectra acquisition and 150 l/mm grating was used with a 100 μm input slit to collect those spectra. 2D CL spectral mapping was performed at 5 kV and 230 pA over a 15 × 11 μm area with a pixel density of 140 × 110 using a 200 ms acquisition time per pixel. All CL measurements were conducted at room temperature.

*Raman/PL.* The PL and Raman maps were obtained using two different setups, both yielding similar results. The first setup utilized the commercially available InVia Qontor system (Renishaw). Excitation was performed with a 532 nm laser, delivered through a 100× (NA = 0.85) Leica objective. The scattered signal was collected by the same objective and passed through a set of ultra-narrow notch filters onto 1800 and 300 lines $mm^{-1}$ gratings for Raman and PL measurements, respectively.

The second setup was a custom-built micro-PL system. PL excitation was carried out using a 532 nm laser (Excelsior, Spectra Physics, 100 mW) through an upright microscope with a 100× objective (NA = 0.9). The emitted PL light was analyzed by a spectrometer (Spectra Pro 2300i, Acton, f = 0.3 m), which was coupled to the microscope and equipped with 150, 600, and 1800 lines $mm^{-1}$ gratings, as well as a CCD camera (Pixis 256BR, Princeton Instruments).

*Device Fabrication.* Regular clean room techniques were used. For Au adhesion, 5nm Ti / Cr were deposited before Au.

*STEM.* Atomic resolution high-angle annular dark field (HAADF)scanning transmission electron microscope (STEM) images of bilayer hBN were acquired using a Nion UltraSTEM 200 operated with an accelerating voltage of 80 kV.

*ODMR.* For continuous-wave (CW) optically detected magnetic resonance (ODMR) characterization, the hBN film was transferred onto a pre-patterned waveguide for microwave (MW) excitation. The MW signal was generated using an OPX+ and Octave system (Quantum Machines), providing a frequency range of 2 GHz–18 GHz. An amplifier was used to deliver 26 dBm of MW power to the system. The spin defect ensemble was optically excited using a 516 nm Cobolt laser with a fluence of around 20 W/cm². The resulting photoluminescence (PL) was collected by a single-photon counting module (SPCM, Excelitas) after passing through a 750 nm long-pass filter. The ODMR contrast was measured by monitoring PL intensity as a function of the swept MW frequency.

*$T_1$ relaxation.* The $T_1$ spin relaxation measurement is performed using an all-optical scheme. The spin state is initially polarized by an optical pulse (516 nm, 5 μs), followed by a second optical pulse (516 nm,

5 µs) for readout. A single photon counting module (SPCM, Excelitas) detects photoluminescence (PL) within a 200 ns window, synchronized with the arrival of the readout pulse. The $T_1$ relaxation time is extracted by measuring the PL intensity as a function of the delay between the initialization and readout pulses and fitting the data to a single exponential decay for time constant analysis.

*DFT calculations*. First-principles DFT calculations were performed using the plane-wave Vienna *Ab initio* Simulation Package (VASP).[49] The electron-ion interactions were described by the projector-augmented-wave (PAW) method. The exchange-correlation interactions were captured by the local density approximation (LDA).[50] For a primitive unit cell of pristine monolayer hBN, atomic positions and lattice constants were optimized until the residual forces were below 0.001 eV/Å and the cutoff energy was chosen at 500 eV. A gamma-centered 36×36×1 k-point sampling was used, and the optimized lattice constants were $a = b = 2.489$ Å. A vacuum region of about 27 Å in the *z* direction was used to avoid spurious interactions with the neighboring cells. Then, a 6×6×1 supercell was constructed where a $V_B^-$ defect or a $N_B V_N$ defect was introduced. The k-point sampling was reduced to 6×6×1 accordingly. Atomic positions were optimized until the residual forces were below 0.001 eV/Å. Phonon calculations were carried out in this optimized supercell using the finite difference scheme implemented in the Phonopy software.[51] Hellmann-Feynman forces in the supercell were computed by VASP for both positive and negative atomic displacements ($\Delta = 0.01$ Å) and then used in Phonopy to construct the dynamic matrix, whose diagonalization provides phonon frequencies and phonon eigenvectors (*i.e.*, vibrations). Raman intensities of the supercell structure were calculated within the Placzek approximation.[52, 53] Basically, one needs to calculate the derivatives of the dielectric tensors with respect to the atomic displacements for obtaining the Raman tensor. For both positive and negative atomic displacements ($\Delta = 0.01$ Å), the dielectric tensors $\varepsilon_{\alpha\beta}$ were computed by VASP at the experimental laser frequency 2.33 eV (532 nm) and thus their derivatives were obtained via the finite difference scheme. Based on the phonon frequencies, phonon eigenvectors and the derivatives of dielectric tensors, Raman tensor $\tilde{R}$ and Raman intensity of any phonon mode can be obtained. Finally, a Raman spectrum was obtained after Lorentzian broadening with a full width at half maximum (FWHM) of 30 cm$^{-1}$.

**Acknowledgements.**

This work was supported by the Center for Nanophase Materials Sciences (CNMS), which is a US Department of Energy, Office of Science User Facility at Oak Ridge National Laboratory (ORNL). The hBN synthesis and spin defect characterization efforts were sponsored by the U.S. Department of Energy, Office of Science, Basic Energy Sciences, Materials Sciences and Engineering Division. The work at the University of Aveiro (Portugal) was developed within the scope of the project CICECO-Aveiro Institute of Materials, UIDB/50011/2020 (DOI: 10.54499/UIDB/50011/2020), UIDP/50011/2020 (DOI: 10.54499/UIDP/50011/2020) & LA/P/0006/2020 (DOI: 10.54499/LA/P/0006/2020), financed by national funds through the FCT/MCTES (PIDDAC). A.T. acknowledges individual support by the 2021.03599.CEECIND/CP1659/CT0016 contract (DOI:10.54499/2021.03599.CEECIND/CP1659/CT0016) through national funds provided by FCT – Fundação para a Ciência e a Tecnologia.

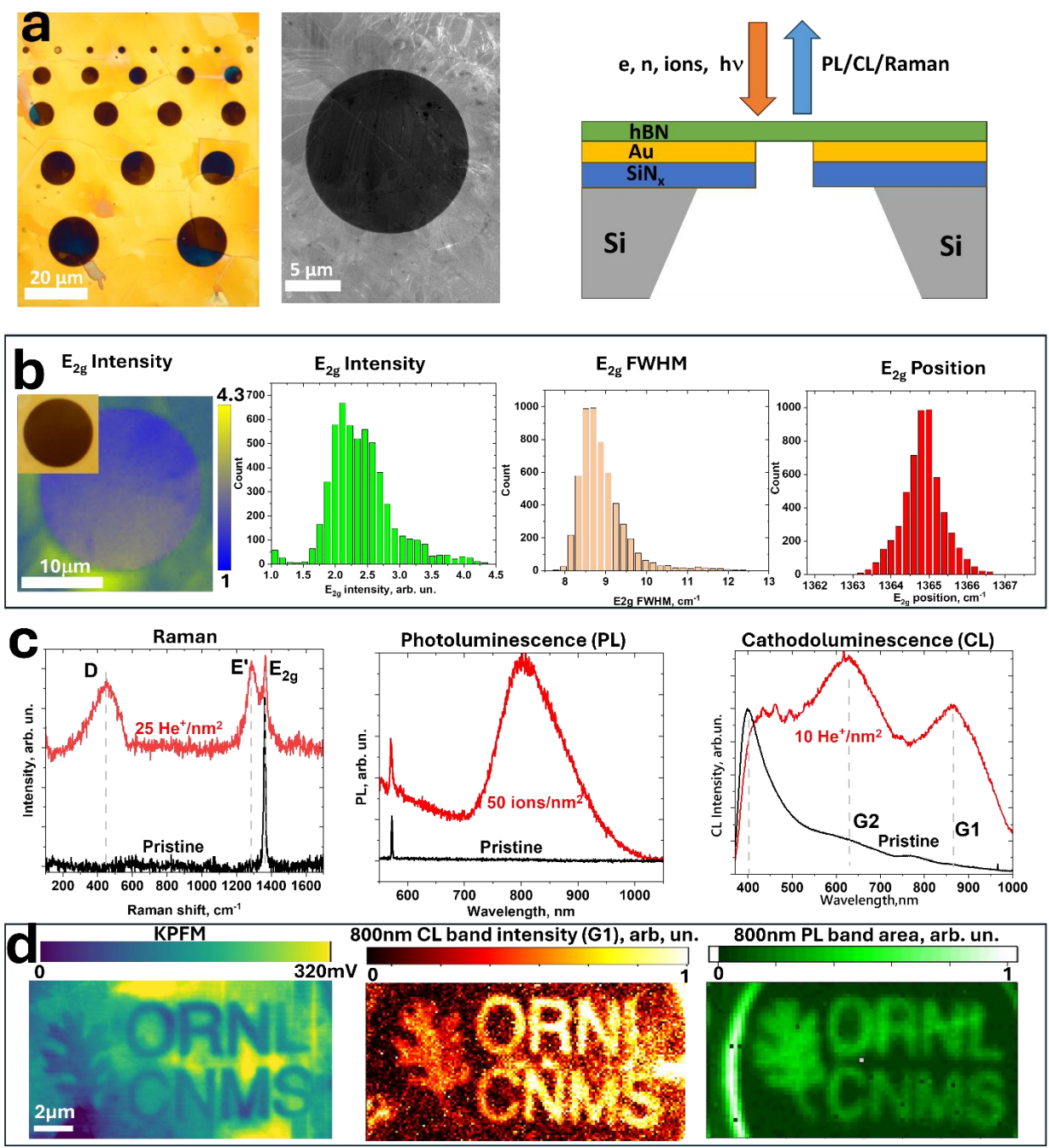

**Figure 1.** *Description of suspended hBN samples and the main spectroscopic features induced by He$^+$ bombardment.* **a.** Optical microscope image of suspended hBN on a Si/SiN$_x$/Au membrane (left). SEM image of the same sample (center) and a schematic of the chip (right). **b.** Characteristic Raman maps of suspended CVD-grown hBN, along with histograms of E$_{2g}$ peak intensity, FWHM and position. **c.** Changes in Raman spectra (left), photoluminescence (PL) (center) and cathodoluminescence (CL) (right) after bombardment by 30keV He$^+$ ions. The ion doses are indicated in the figure. **c.** Maps of suspended, patterned hBN membranes with L ~ 120 nm thickness obtained by KPFM (left), CL (center) and PL (right).

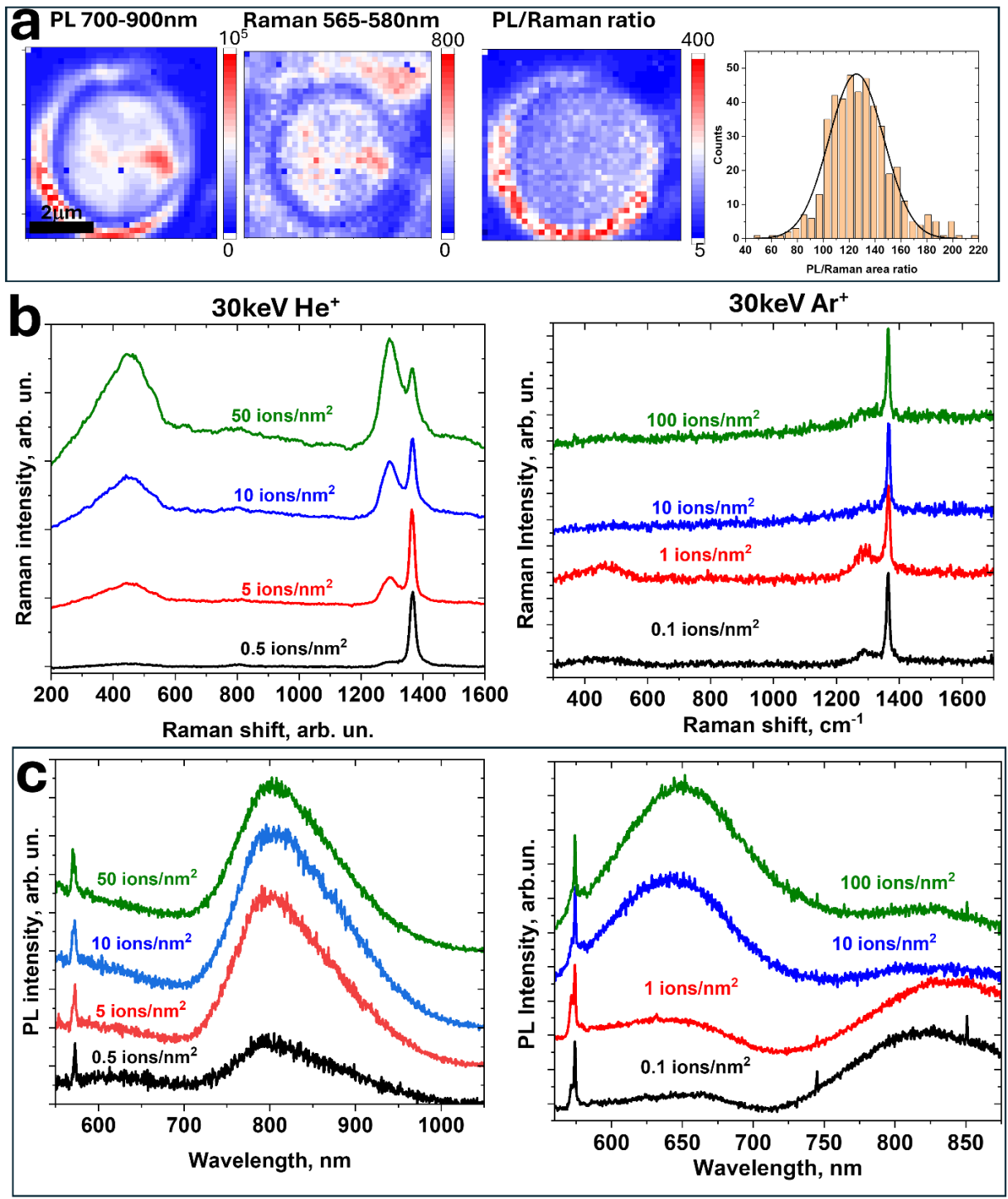

**Figure 2.** *PL signal @800 nm normalized to hBN thickness and spectroscopic variations in suspended hBN under bombardment by various ions and dosages.* **a.** Maps of PL @800 nm (left) and Raman signal for He[+] irradiated suspended sample for the 50 ions/nm$^2$ dosage, along with the PL/Raman ratio and its histogram over the suspended region (right) for. **b.** Raman spectra of suspended hBN for He[+] (left) and Ar[+] (right) 30 keV bombardment. **c.** Same as (b), but for PL spectra.

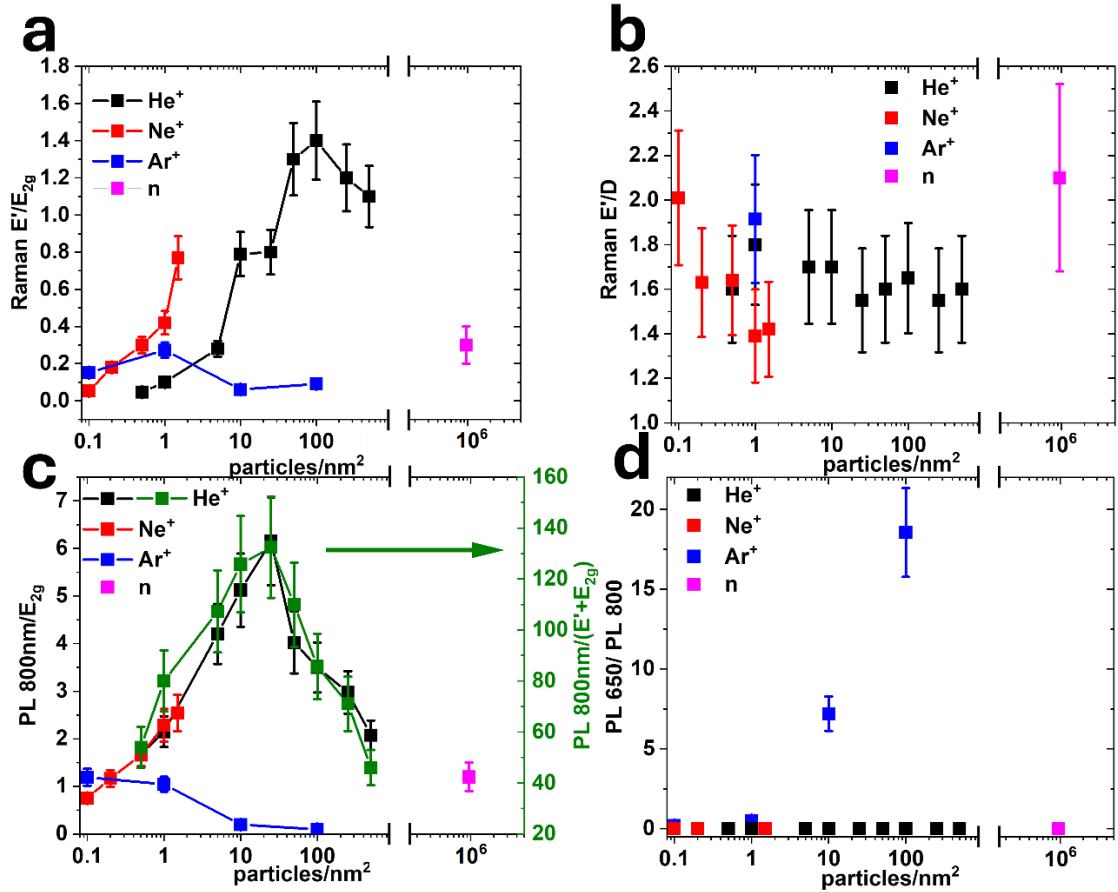

**Figure 3.** *Spectroscopic assessment of suspended hBN under bombardment by different ions, $He^+$ (black), $Ne^+$ (red) and $Ar^+$ (green) and neutrons (magenta).* **a.** Intensity ratio of Raman bands E' to $E_{2g}$. **b.** Intensity ratio of Raman bands E'/D. **c.** PL intensity at 800nm normalized by $E_{2g}$ Raman signal; area ratios for $He^+$ are highlighted in green. **d.** Intensity ratio of PL bands at 800 nm and 650 nm.

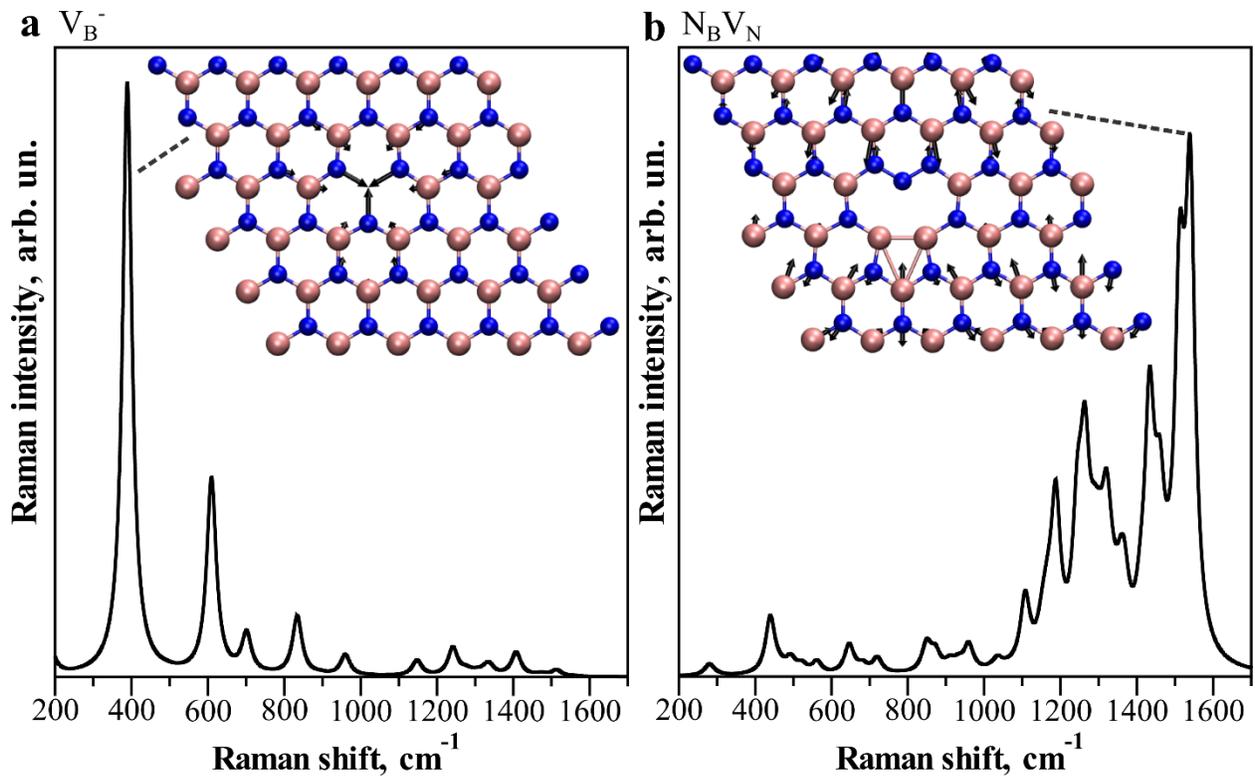

**Figure 4.** *Calculated Raman spectra of hBN with different defects.* **a.** $V_B^-$ defect and **b.** $N_BV_N$ defect. The atomic vibration patterns of dominant Raman peaks are shown as insets. Black arrows indicate the direction and amplitude of the atomic vibrations. B atoms are in pink color and N atoms in blue color

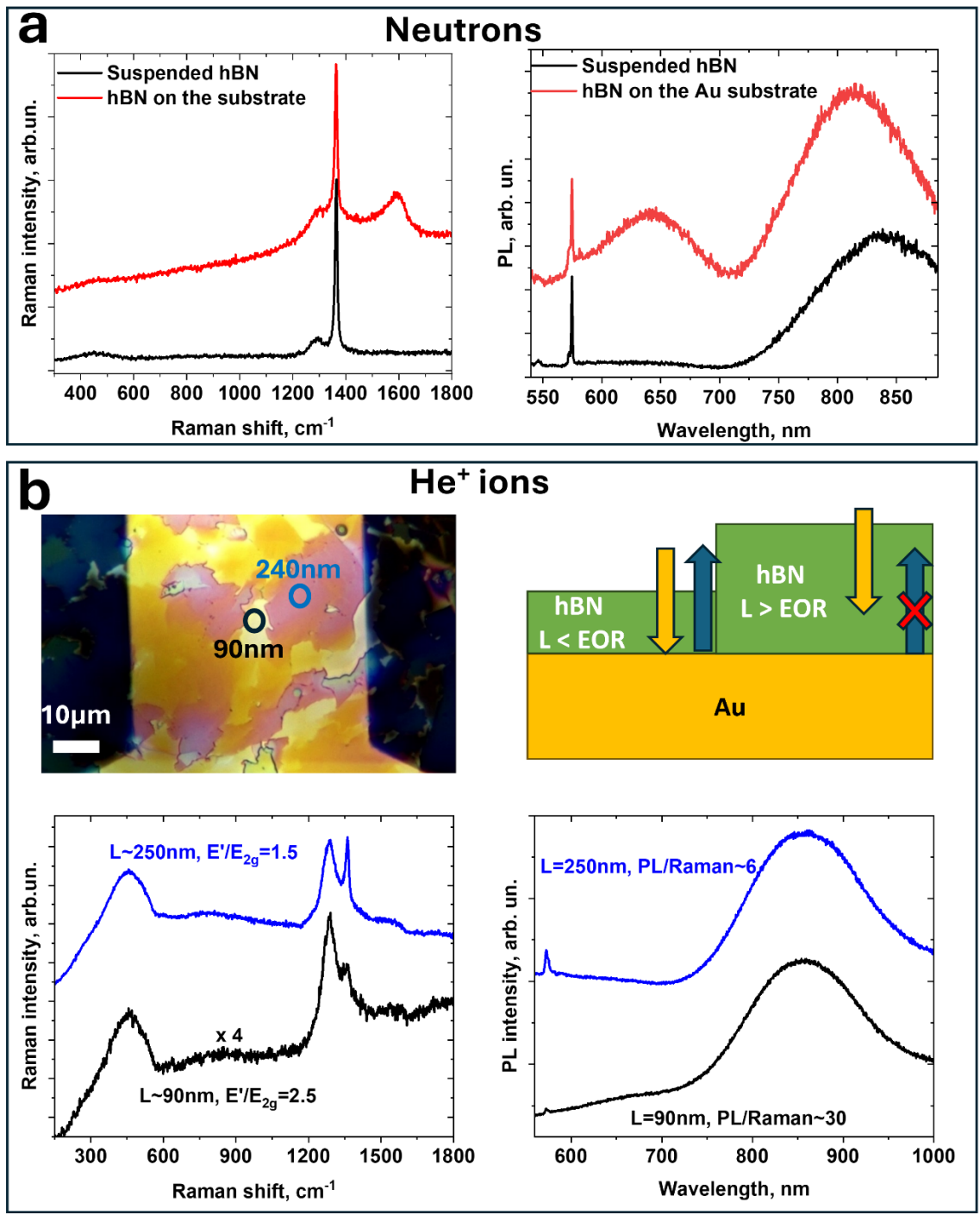

**Figure 5.** *Effects of substrate on defect creation.* **a.** Raman (left) and PL spectra (right) of suspended and supported hBN by the substrate. Note emergence of 1580 cm$^{-1}$ band in Raman and 650nm PL for supported samples. **b.** He$^+$ ion bombarded hBN on a gold substrate. 50 He$^+$/nm$^2$ example. Optical microscope image of a non-uniform hBN transferred onto a gold substrate with approximate hBN thickness indicated at two points, each marked by a different color (top). Raman and PL spectra corresponding to these two points show characteristics that differ significantly from those observed in suspended samples, emphasizing the importance of considering the effects of substrate (bottom).

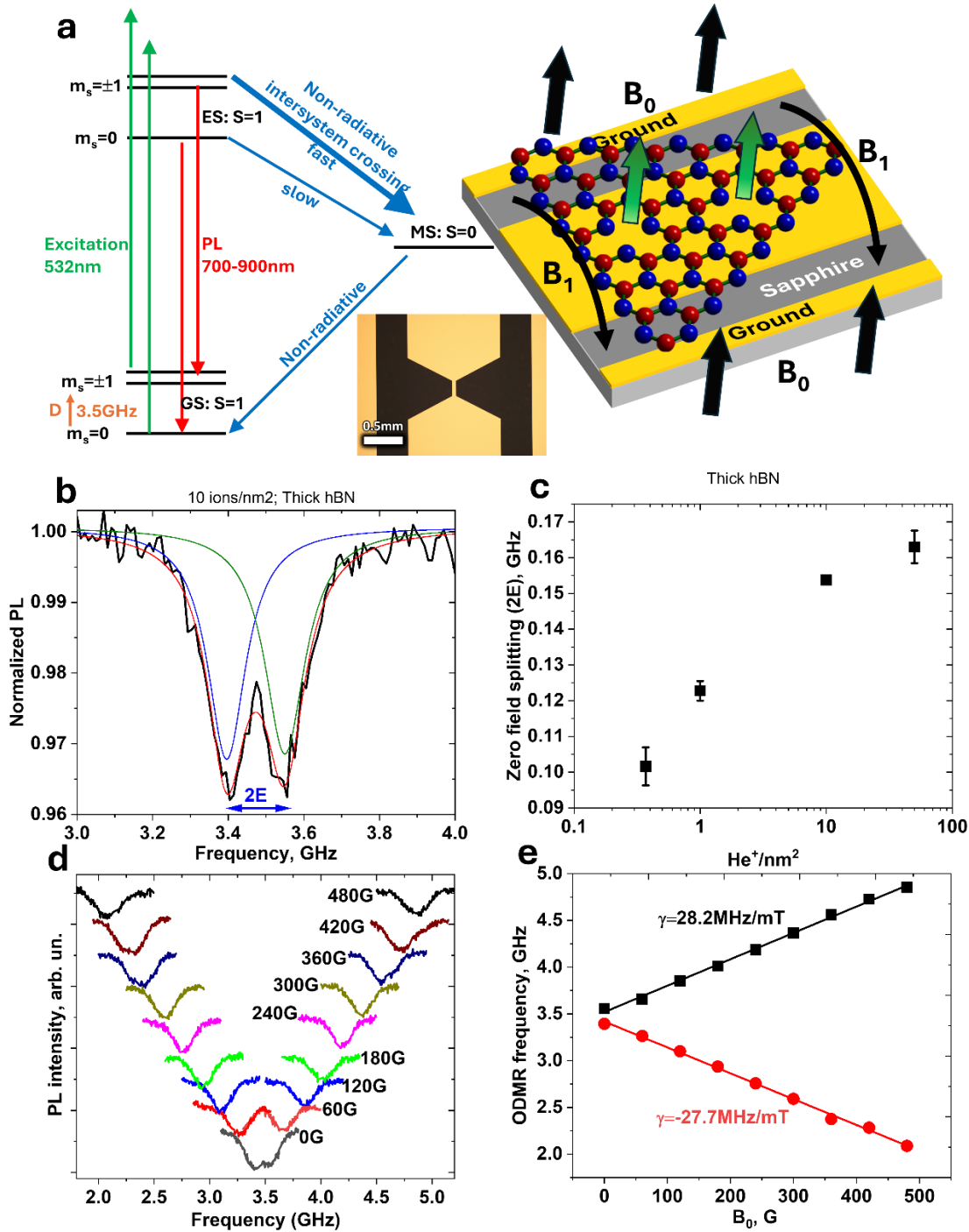

**Figure 6.** *Optically detected magnetic resonance (ODMR) of created defect.* ***a.*** Simplified energy diagram of the $V_B^-$ defect (left) and a schematic of experimental geometry (right), along with an optical microscope image of the coplanar waveguide design (insert). ***b.*** Characteristic ODMR spectrum at zero magnetic field ($B_0$). ***c.*** Dependance of zero-field splitting (ZFS) parameter *E* on the $He^+$ bombardment dosage. The parameter *D* remains unaffected by bombardment ***d.*** ODMR spectra at varying $B_0$. ***e.*** Dependence of $\nu_1$ and $\nu_2$ on the $B_0$ yielding g~2.00.

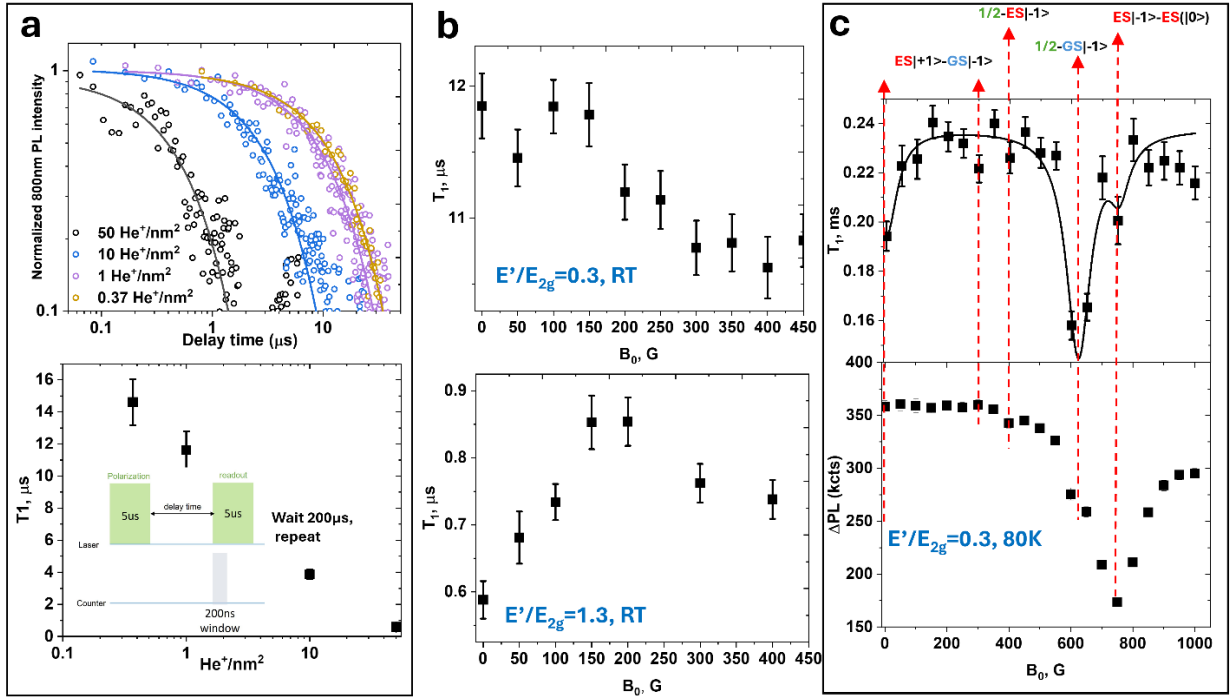

**Figure 7.** *Dependence of the spin-lattice relaxation time ($T_1$) on the defect density and applied magnetic field.* **a.** Time traces used for $T_1$ estimation (top) and $T_1$ time vs. He$^+$ ion dosages (bottom). **b.** Room temperature $T_1$ as a function of the applied magnetic field for low $V_B^-$ defect densities (E'/E$_{2g}$=0.3, top) and high defect densities (E'/E$_{2g}$=1.3, bottom). **c** $T_1$ as a function of the applied magnetic field for low $V_B^-$ defect densities at T=80K (top) and ΔPL, a PL decay amplitude (bottom); the time trace was fitted as y0+ΔPL*exp(-t/$T_1$).

**References.**

(1) Wolfowicz, G.; Heremans, F. J.; Anderson, C. P.; Kanai, S.; Seo, H.; Gali, A.; Galli, G.; Awschalom, D. D. Quantum guidelines for solid-state spin defects (Oct, 10.1038/s41578-021-00306-y, 2021). *Nat Rev Mater* **2021**, *6* (12), 1191-1191. DOI: 10.1038/s41578-021-00393-x.
(2) Aharonovich, I.; Englund, D.; Toth, M. Solid-state single-photon emitters. *Nat Photonics* **2016**, *10* (10), 631-641. DOI: 10.1038/Nphoton.2016.186.
(3) Fang, H. H.; Wang, X. J.; Marie, X.; Sun, H. B. Quantum sensing with optically accessible spin defects in van der Waals layered materials. *Light-Sci Appl* **2024**, *13* (1). DOI: ARTN 303 10.1038/s41377-024-01630-y.

(4) Hong, J. H.; Jin, C. H.; Yuan, J.; Zhang, Z. Atomic Defects in Two-Dimensional Materials: From Single-Atom Spectroscopy to Functionalities in Opto-/Electronics, Nanomagnetism, and Catalysis. *Adv Mater* **2017**, *29* (14). DOI: ARTN 1606434 10.1002/adma.201606434.

(5) Vaidya, S.; Gao, X. Y.; Dikshit, S.; Aharonovich, I.; Li, T. C. Quantum sensing and imaging with spin defects in hexagonal boron nitride. *Adv Phys-X* **2023**, *8* (1). DOI: Artn 2206049

10.1080/23746149.2023.2206049.
(6) Tran, T. T.; Bray, K.; Ford, M. J.; Toth, M.; Aharonovich, I. Quantum emission from hexagonal boron nitride monolayers. *Nat Nanotechnol* **2016**, *11* (1), 37-+. DOI: 10.1038/Nnano.2015.242.
(7) Grosso, G.; Moon, H.; Lienhard, B.; Ali, S.; Efetov, D. K.; Furchi, M. M.; Jarillo-Herrero, P.; Ford, M. J.; Aharonovich, I.; Englund, D. Tunable and high-purity room temperature single-photon emission from atomic defects in hexagonal boron nitride. *Nat Commun* **2017**, *8*. DOI: ARTN 705 10.1038/s41467-017-00810-2.

(8) Mendelson, N.; Chugh, D.; Reimers, J. R.; Cheng, T. S.; Gottscholl, A.; Long, H.; Mellor, C. J.; Zettl, A.; Dyakonov, V.; Beton, P. H.; et al. Identifying carbon as the source of visible single-photon emission from hexagonal boron nitride. *Nat Mater* **2021**, *20* (3), 321-+. DOI: 10.1038/s41563-020-00850-y.
(9) Tran, T. T.; Elbadawi, C.; Totonjian, D.; Lobo, C. J.; Grosso, G.; Moon, H.; Englund, D. R.; Ford, M. J.; Aharonovich, I.; Toth, M. Robust Multicolor Single Photon Emission from Point Defects in Hexagonal Boron Nitride. *Acs Nano* **2016**, *10* (8), 7331-7338. DOI: 10.1021/acsnano.6b03602.
(10) Gottscholl, A.; Kianinia, M.; Soltamov, V.; Orlinskii, S.; Mamin, G.; Bradac, C.; Kasper, C.; Krambrock, K.; Sperlich, A.; Toth, M.; et al. Initialization and read-out of intrinsic spin defects in a van der Waals crystal at room temperature. *Nat Mater* **2020**, *19* (5), 540-+. DOI: 10.1038/s41563-020-0619-6.
(11) Venturi, G.; Chiodini, S.; Melchioni, N.; Janzen, E.; Edgar, J. H.; Ronning, C.; Ambrosio, A. Selective Generation of Luminescent Defects in Hexagonal Boron Nitride. *Laser Photonics Rev* **2024**, *18* (6). DOI: 10.1002/lpor.202300973.
(12) al., T. e. Structured-Defect Engineering of Hexagonal Boron Nitride for Identified Visible Single-Photon Emitters. *Acs Nano ASAP*.
(13) Kabyshev, A. V.; Kezkalo, V. M.; Lopatin, V. V.; Serikov, L. V.; Surov, Y. P.; Shiyan, L. N. Postradiation Defects in Neutron-Irradiated Pyrolytic Boron-Nitride. *Physica Status Solidi a-Applied Research* **1991**, *126* (1), K19-K23. DOI: DOI 10.1002/pssa.2211260136.
(14) Gottscholl, A.; Diez, M.; Soltamov, V.; Kasper, C.; Sperlich, A.; Kianinia, M.; Bradac, C.; Aharonovich, I.; Dyakonov, V. Room temperature coherent control of spin defects in hexagonal boron nitride. *Sci Adv* **2021**, *7* (14). DOI: ARTN eabf3630 10.1126/sciadv.abf3630.

(15) Sarkar, S.; Xu, Y.; Mathew, S.; Lal, M.; Chung, J. Y.; Lee, H. Y.; Watanabe, K.; Taniguchi, T.; Venkatesan, T.; Gradecak, S. Identifying Luminescent Boron Vacancies in h-BN Generated Using Controlled He Ion Irradiation. *Nano Lett* **2023**, *24* (1), 43-50. DOI: 10.1021/acs.nanolett.3c03113.

(16) Baber, S.; Malein, R. N. E.; Khatri, P.; Keatley, P. S.; Guo, S.; Withers, F.; Ramsay, A. J.; Luxmoore, I. J. Excited State Spectroscopy of Boron Vacancy Defects in Hexagonal Boron Nitride Using Time-Resolved Optically Detected Magnetic Resonance. *Nano Lett* **2022**, *22* (1), 461-467. DOI: 10.1021/acs.nanolett.1c04366.

(17) Das, S.; Melendez, A. L.; Kao, I. H.; García-Monge, J. A.; Russell, D.; Li, J. H.; Watanabe, K.; Taniguchi, T.; Edgar, J. H.; Katoch, J.; et al. Quantum Sensing of Spin Dynamics Using Boron-Vacancy Centers in Hexagonal Boron Nitride. *Phys Rev Lett* **2024**, *133* (16). DOI: ARTN 166704 10.1103/PhysRevLett.133.166704.

(18) Rizzato, R.; Schalk, M.; Mohr, S.; Hermann, J. C.; Leibold, J. P.; Bruckmaier, F.; Salvitti, G.; Qian, C. J.; Ji, P. R.; Astakhov, G. V.; et al. Extending the coherence of spin defects in hBN enables advanced qubit control and quantum sensing. *Nat Commun* **2023**, *14* (1). DOI: ARTN 5089 10.1038/s41467-023-40473-w.

(19) Liu, W.; Ivády, V.; Li, Z. P.; Yang, Y. Z.; Yu, S.; Meng, Y.; Wang, Z. A.; Guo, N. J.; Yan, F. F.; Li, Q.; et al. Coherent dynamics of multi-spin $V_B^-$ center in hexagonal boron nitride (vol 13, 5713, 2022). *Nat Commun* **2023**, *14* (1). DOI: ARTN 3519 10.1038/s41467-023-39331-6.

(20) Hennessey. Framework for Engineering of Spin Defects in HexagonalBoron Nitride by Focused Ion Beams. *Adv. Quantum Technol.* **2025**. DOI: 10.1002/qute.202300459.

(21) Vlassiouk, I.; Smirnov, S.; Puretzky, A.; Olunloyo, O.; Geohegan, D. B.; Dyck, O.; Lupini, A. R.; Unocic, R. R.; Meyer, H. I. I. I.; Xiao, K.; et al. Armor for Steel: Facile Synthesis of Hexagonal Boron Nitride Films on Various Substrates (Adv. Mater. Interfaces 1/2024). *Adv Mater Interfaces* **2024**, *11* (1). DOI: 10.1002/admi.202470001.

(22) Shi, Z. Y.; Wang, X. J.; Li, Q. T.; Yang, P.; Lu, G. Y.; Jiang, R.; Wang, H. S.; Zhang, C.; Cong, C. X.; Liu, Z.; et al. Vapor-liquid-solid growth of large-area multilayer hexagonal boron nitride on dielectric substrates. *Nat Commun* **2020**, *11* (1). DOI: ARTN 849 10.1038/s41467-020-14596-3.

(23) Fukamachi, S.; Solis-Fernandez, P.; Kawahara, K.; Tanaka, D.; Otake, T.; Lin, Y. C.; Suenaga, K.; Ago, H. Large-area synthesis and transfer of multilayer hexagonal boron nitride for enhanced graphene device arrays. *Nat Electron* **2023**, *6* (2), 126-+. DOI: 10.1038/s41928-022-00911-x.

(24) Naclerio, A. E.; Kidambi, P. R. A Review of Scalable Hexagonal Boron Nitride (h-BN) Synthesis for Present and Future Applications. *Adv Mater* **2023**, *35* (6). DOI: 10.1002/adma.202207374.

(25) Li, J. H.; Glaser, E. R.; Elias, C.; Ye, G. H.; Evans, D.; Xue, L. J.; Liu, S.; Cassabois, G.; Gil, B.; Valvin, P.; et al. Defect Engineering of Monoisotopic Hexagonal Boron Nitride Crystals Neutron Transmutation Doping. *Chem Mater* **2021**, *33* (23), 9231-9239. DOI: 10.1021/acs.chemmater.1c02849.

(26) Linderälv, C.; Wieczorek, W.; Erhart, P. Vibrational signatures for the identification of single-photon emitters in hexagonal boron nitride. *Phys Rev B* **2021**, *103* (11). DOI: ARTN 115421 10.1103/PhysRevB.103.115421.

(27) Li, Y. S.; Xie, X. H.; Zeng, H.; Li, B. H.; Zhang, Z. Z.; Wang, S. P.; Liu, J. S.; Shen, D. Z. Giant moire trapping of excitons in twisted hBN. *Opt Express* **2022**, *30* (7), 10596-10604. DOI: 10.1364/Oe.450409.

(28) Bianco, F.; Corte, E.; Tchernij, S. D.; Forneris, J.; Fabbri, F. Engineering Multicolor Radiative Centers in hBN Flakes by Varying the Electron Beam Irradiation Parameters. *Nanomaterials-Basel* **2023**, *13* (4). DOI: ARTN 739 10.3390/nano13040739.

(29) Kumar, A.; Cholsuk, C.; Mishuk, M. N.; Hazra, M.; Pillot, C.; Matthes, T.; Shaik, T. A.; Çakan, A.; Deckert, V.; Suwanna, S.; et al. Comparative Study of Quantum Emitter Fabrication in Wide Bandgap Materials Using Localized Electron Irradiation. *Acs Appl Opt Mater* **2024**, *2* (2), 323-332. DOI: 10.1021/acsaom.3c00441.

(30) Fournier, C.; Plaud, A.; Roux, S.; Pierret, A.; Rosticher, M.; Watanabe, K.; Taniguchi, T.; Buil, S.; Quélin, X.; Barjon, J.; et al. Position-controlled quantum emitters with reproducible emission wavelength in hexagonal boron nitride. *Nat Commun* **2021**, *12* (1). DOI: ARTN 3779 10.1038/s41467-021-24019-6.

(31) Nedic, S.; Yamamura, K.; Gale, A.; Aharonovich, I.; Toth, M. Electron Beam Restructuring of Quantum Emitters in Hexagonal Boron Nitride. *Adv Opt Mater* **2024**, *12* (24). DOI: 10.1002/adom.202400908.

(32) Gale, A.; Li, C.; Chen, Y. L.; Watanabe, K.; Taniguchi, T.; Aharonovich, I.; Toth, M. Site-Specific Fabrication of Blue Quantum Emitters in Hexagonal Boron Nitride. *Acs Photonics* **2022**, *9* (6), 2170-2177. DOI: 10.1021/acsphotonics.2c00631.

(33) Silly, M. G.; Jaffrennou, P.; Barjon, J.; Lauret, J. S.; Ducastelle, F.; Loiseau, A.; Obraztsova, E.; Attal-Tretout, B.; Rosencher, E. Luminescence properties of hexagonal boron nitride: Cathodoluminescence and photoluminescence spectroscopy measurements. *Phys Rev B* **2007**, *75* (8). DOI: ARTN 085205 10.1103/PhysRevB.75.085205.

(34) Chaturvedi, P.; Vlassiouk, I. V.; Cullen, D. A.; Rondinone, A. J.; Lavrik, N. V.; Smirnov, S. N. Ionic Conductance through Graphene: Assessing Its Applicability as a Proton Selective Membrane. *Acs Nano* **2019**, *13* (10), 12109-12119. DOI: 10.1021/acsnano.9b06505.

(35) Lau, D.; Hughes, A. E.; Muster, T. H.; Davis, T. J.; Glenn, A. M. Electron-Beam-Induced Carbon Contamination on Silicon: Characterization Using Raman Spectroscopy and Atomic Force Microscopy. *Microsc Microanal* **2010**, *16* (1), 13-20. DOI: 10.1017/S1431927609991206.

(36) de Vera, P.; Garcia-Molina, R. Electron Inelastic Mean Free Paths in Condensed Matter Down to a Few Electronvolts. *J Phys Chem C* **2019**, *123* (4), 2075-2083. DOI: 10.1021/acs.jpcc.8b10832.

(37) Tanuma, S.; Shiratori, T.; Kimura, T.; Goto, K.; Ichimura, S.; Powell, C. J. Experimental determination of electron inelastic mean free paths in 13 elemental solids in the 50 to 5000 eV energy range by elastic-peak electron spectroscopy. *Surf Interface Anal* **2005**, *37* (11), 833-845. DOI: 10.1002/sia.2102.

(38) Eckmann, A.; Felten, A.; Mishchenko, A.; Britnell, L.; Krupke, R.; Novoselov, K. S.; Casiraghi, C. Probing the Nature of Defects in Graphene by Raman Spectroscopy. *Nano Lett* **2012**, *12* (8), 3925-3930. DOI: 10.1021/nl300901a.

(39) Krivanek, O. L.; Chisholm, M. F.; Nicolosi, V.; Pennycook, T. J.; Corbin, G. J.; Dellby, N.; Murfitt, M. F.; Own, C. S.; Szilagyi, Z. S.; Oxley, M. P.; et al. Atom-by-atom structural and chemical analysis by annular dark-field electron microscopy. *Nature* **2010**, *464* (7288), 571-574. DOI: 10.1038/nature08879.

(40) F.I., B. D. O. a. A. Atomic Engineering of Triangular Nanopores in Monolayer hBN for Membrane Applications: A Decoupled Seeding and Growth Approach. *ACS Appl. Nano Mater* **2025**, *ASAP*. DOI: https://doi.org/10.1021/acsanm.4c06998.

(41) Cohen-Tanugi, D.; Yao, N. Superior imaging resolution in scanning helium-ion microscopy: A look at beam-sample interactions. *J Appl Phys* **2008**, *104* (6). DOI: Artn 063504

10.1063/1.2976299.

(42) Qi, Y. P.; Sadi, M. A.; Hu, D.; Zheng, M.; Wu, Z. P.; Jiang, Y. C.; Chen, Y. P. Recent Progress in Strain Engineering on Van der Waals 2D Materials: Tunable Electrical, Electrochemical, Magnetic, and Optical Properties. *Adv Mater* **2023**, *35* (12). DOI: 10.1002/adma.202205714.

(43) Curie, D.; Krogel, J. T.; Cavar, L.; Solanki, A.; Upadhyaya, P.; Li, T. C.; Pai, Y. Y.; Chilcote, M.; Iyer, V.; Puretzky, A.; et al. Correlative Nanoscale Imaging of Strained hBN Spin Defects. *Acs Appl Mater Inter* **2022**. DOI: 10.1021/acsami.2c11886.

(44) Mathur, N.; Mukherjee, A.; Gao, X. Y.; Luo, J. L.; McCullian, B. A.; Li, T. C.; Vamivakas, A. N.; Fuchs, G. D. Excited-state spin-resonance spectroscopy of $V_B^-$ defect centers in hexagonal boron nitride. *Nat Commun* **2022**, *13* (1). DOI: ARTN 3233

10.1038/s41467-022-30772-z.

(45) Jarmola, A.; Berzins, A.; Smits, J.; Smits, K.; Prikulis, J.; Gahbauer, F.; Ferber, R.; Erts, D.; Auzinsh, M.; Budker, D. Longitudinal spin-relaxation in nitrogen-vacancy centers in electron irradiated diamond. *Appl Phys Lett* **2015**, *107* (24). DOI: Artn 242403

10.1063/1.4937489.

(46) Gong, R. T.; He, G. H.; Gao, X. Y.; Ju, P.; Liu, Z. Y.; Ye, B. T.; Henriksen, E. A.; Li, T. C.; Zu, C. Coherent dynamics of strongly interacting electronic spin defects in hexagonal boron nitride. *Nat Commun* **2023**, *14* (1). DOI: ARTN 3299

10.1038/s41467-023-39115-y.

(47) Jarmola, A.; Acosta, V. M.; Jensen, K.; Chemerisov, S.; Budker, D. Temperature- and Magnetic-Field-Dependent Longitudinal Spin Relaxation in Nitrogen-Vacancy Ensembles in Diamond. *Phys Rev Lett* **2012**, *108* (19). DOI: ARTN 197601

10.1103/PhysRevLett.108.197601.

(48) Wang, H. J.; Shin, C. S.; Seltzer, S. J.; Avalos, C. E.; Pines, A.; Bajaj, V. S. Optically detected cross-relaxation spectroscopy of electron spins in diamond. *Nat Commun* **2014**, *5*. DOI: ARTN 4135

10.1038/ncomms5135.

(49) Kresse, G.; Furthmuller, J. Efficiency of ab-initio total energy calculations for metals and semiconductors using a plane-wave basis set. *Comp Mater Sci* **1996**, *6* (1), 15-50. DOI: Doi 10.1016/0927-0256(96)00008-0.

(50) Kresse, G.; Furthmuller, J. Efficient iterative schemes for ab initio total-energy calculations using a plane-wave basis set. *Phys Rev B* **1996**, *54* (16), 11169-11186. DOI: DOI 10.1103/PhysRevB.54.11169.

(51) Togo, A.; Oba, F.; Tanaka, I. First-principles calculations of the ferroelastic transition between rutile-type and CaCl-type SiO at high pressures. *Phys Rev B* **2008**, *78* (13). DOI: ARTN 134106

10.1103/PhysRevB.78.134106.

(52) Umari, P.; Pasquarello, A.; Dal Corso, A. Raman scattering intensities in α-quartz:: A first-principles investigation -: art. no. 094305. *Phys Rev B* **2001**, *63* (9). DOI: ARTN 094305

DOI 10.1103/PhysRevB.63.094305.

(53) Kong, X. R.; Ganesh, P.; Liang, L. B. First-principles study of the magneto-Raman effect in van der Waals layered magnets. *Npj 2d Mater Appl* **2024**, *8* (1). DOI: ARTN 82 10.1038/s41699-024-00515-3.